\documentclass{article}
\usepackage{spconf,amsmath,graphicx}
\usepackage{url}

\title{A practical convolutional neural network as loop filter for intra frame}
\name{Xiaodan Song$^*$, Jiabao Yao$^*$, Lulu Zhou$^*$, Li Wang, Xiaoyang Wu, Di Xie and Shiliang Pu\thanks{$^*$ Equal contribution}}
\address{$\{$songxiaodan,yaojiabao,zhoululu,wangli7,wuxiaoyang,xiedi,pushiliang$\}$@hikvision.com\\Hikvision Research Institute, Hangzhou, China}
\begin{document}
\topmargin=0mm 
%
\maketitle
\begin{abstract}
Loop filters are used in video coding to remove artifacts or improve performance. Recent advances in deploying convolutional neural network (CNN) to replace traditional loop filters show large gains but with problems for practical application. First, different model is used for frames encoded with different quantization parameter (QP), respectively. It is expensive for hardware. Second, float points operation in CNN leads to inconsistency between encoding and decoding across different platforms. Third, redundancy within CNN model consumes  precious computational resources.

This paper proposes a CNN as the loop filter for intra frames and proposes a scheme to solve the above problems. It aims to design a single CNN model with low redundancy to adapt to decoded frames with different qualities and ensure consistency. To adapt to reconstructions with different qualities, both reconstruction and QP are taken as inputs. After training, the obtained model is compressed to reduce redundancy. To ensure consistency, dynamic fixed points (DFP) are adopted in testing CNN. Parameters in the compressed model are first quantized to DFP and then used for inference of CNN. Outputs of each layer in CNN are computed by DFP operations. Experimental results on JEM 7.0 report 3.14\%, 5.21\%, 6.28\% BD-rate savings for luma and two chroma components with all intra configuration when replacing all traditional filters.  
\end{abstract}
\section{Introduction}
\label{sec:intro}

\begin{figure}[!t]
	\centering
	\vspace{-1em}
	\includegraphics[width=0.8\linewidth]{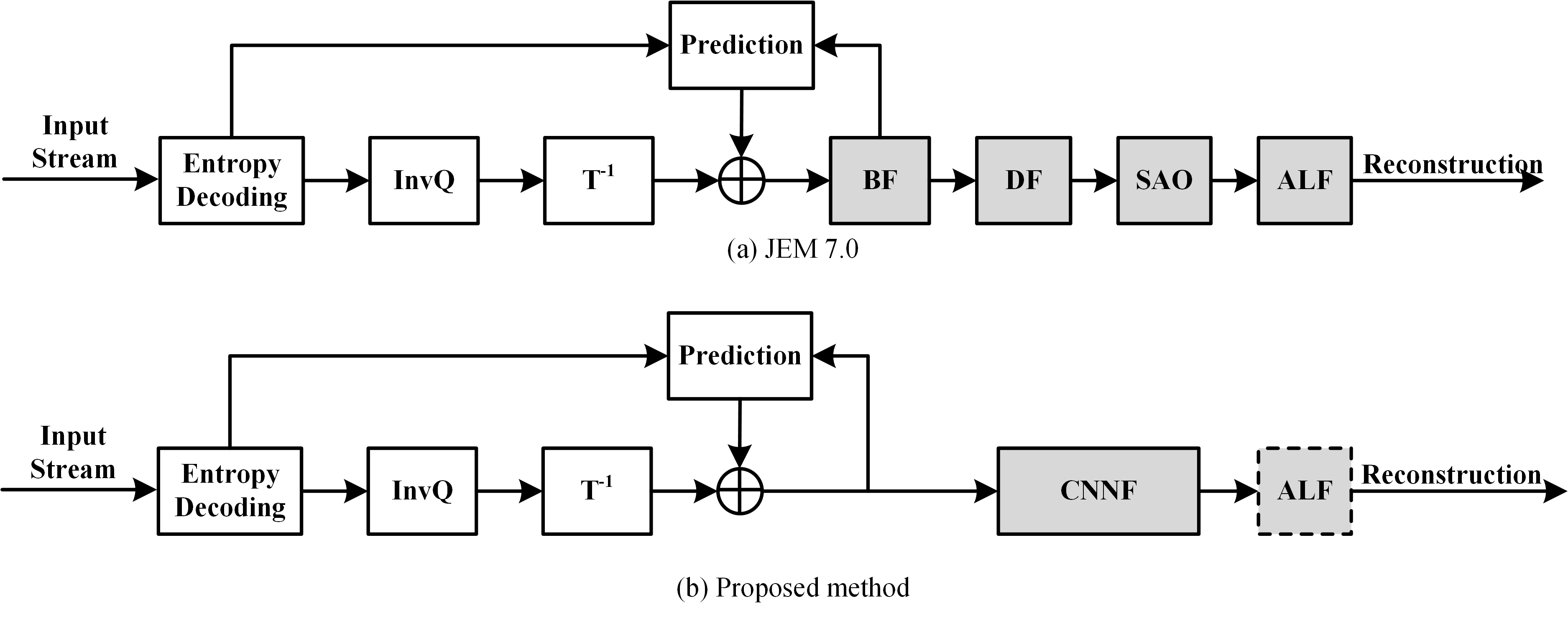}
	\caption{Comparison of intra decoding scheme between JEM 7.0 and the proposed method.}
	\label{framework}
	\vspace{-1.5em}
\end{figure}

In lossy image/video coding, loop filters are usually used to remove artifacts or further improve coding performance. For example, the recent standardized HEVC \cite{sullivan2012overview}, employs sample adaptive offset (SAO) and deblocking filter (DF). Up to four filters are introduced in the reference software JEM 7.0, which is used within the joint video exploration team (JVET) group \cite{JVET} for the next generation of video coding standard. Two additional filters, i.e., bilateral filter (BF) and adaptive loop filter (ALF) are included as shown in Fig. \ref{framework}(a). It is wondering whether they are sufficient to deal with the complex content in nature video and whether they can be replaced by a single type of filter. The recent advances in deep learning have shed lights on this.

Recent research has investigated the deep learning approach, especially convolutional neural network (CNN), on post-processing \cite{VRCNN,wang2017novel, Kang2017MultimodalmultiscaleCN,yang2017decoder} or loop filtering \cite{park2016cnn, jia2017spatial}. In post-processing, CNN are used to improve the subjective or objective quality of reconstructed frames after decoding. It does not require to change the encoding algorithm. While in loop filtering, the filtered reconstruction is used as reference of the following frames and helps to reduce bit-rate.

In post-processing, Dai \emph{et al.} \cite{VRCNN} adopts a CNN with residual learning structure with variable filter size for acceleration . \cite{wang2017novel} designs and trains a deeper network for intra frame and directly deploys it to B and P frames. Yang \emph{et al.}\cite{yang2017decoder} argues that models obtained for intra frames is not good enough for B and P frames and proposes a scalable CNN for devices with different computational resources. To improve the performance, \cite{Kang2017MultimodalmultiscaleCN} introduces a noval CNN taking into multi-scale and partition of coding tree unit into account. In loop filtering, Park \emph{et al.} \cite{park2016cnn} propose to use CNN as in-loop filter to replace DF and SAO in HEVC and reports bit-rate reduction besides objective quality enhancement. However, the generalizability cannot be ensured due to its test data are included in the training set. \cite{jia2017spatial} takes both the current reconstructed block and the co-located block in the nearest reference frame as inputs to jointly exploit the spatial and temporal information.

The above mentioned CNN approaches show significant gains over traditional methods, which makes them attractive to be used in practice. However, three problems impede the way. First, almost all of them train a separate model for each QP. For a codec, in which QP varies from M to N, N-M+1 models are necessary to be stored. It is expensive for a hardware-oriented codec. Second, the default float point operations in CNN computation will lead to inconsistency between encoding and decoding across different platforms, e.g. CPU and GPU, which cannot be acceptable in video communications among various manufactures and users. Third, there is redundancy among parameters in the pre-trained model \cite{denil2013predicting}, which consumes unnecessary storage and computational resources.

In this paper, we propose a practical convolutional neural network filter (CNNF) to replace all traditional filters as shown in Fig. \ref{framework}(b) and propose a scheme to solve the above problems. Both decoded frame and QP are taken as inputs to CNNF to obtain a QP independent model and adapt to reconstructions with different qualities. After training, the obtained model is compressed for acceleration. To ensure the consistency, the compressed model is quantized to DFP \cite{gysel2016hardware} and outputs of each layer are computed by DFP operations during inference. 

Note that CNNF has been submitted to JVET meeting in Gwangju \cite{ours} and an Adhoc Group is set up to investigate deep learning for video compression.

The rest of this paper is organized as follows. Section \ref{sec:overview} overview the proposed CNNF. Section \ref{sec:pagestyle} explains it in detail. And the training process is shown in Section \ref{training}. The experimental results are given in Section \ref{sec:experiments}. Section \ref {sec:conclusion} concludes this paper.
\section{Overview of the proposed scheme}
\label{sec:overview}

The proposed scheme mainly includes three parts: a novel network, model compression and DFP-based inference of CNNF. A fully CNN with residual learning structure is adopted in CNNF. To obtain one QP independent model, a QP map is generated and taken as an input of CNN. After training, the model is compressed by reducing the filter number of each convolutional layer for acceleration. In the compression, two steps are included. First, additional regularization is included in the loss function to help the compression. During training, filters are automatically pruned based on the absolute value of the scale parameter in its corresponding BN layer. Second, the obtained model in the first step is further compressed by low rank approximation with fine-tuning. After that, filters are reconstructed using a much lower basis from the low-rank space for acceleration.

Before DFP inference, parameters in the compressed model are first quantized and converted to DFP. To recover performance loss due to quantization, fine-tuning is established \cite{LinTA15}. During inference of CNNF, outputs of each layer are computed by DFP operations and then quantized to DFP with low bit-width to avoid overflow. After the fixed-point inference, the outputs are denormalized to obtain the final reconstruction.

\begin{figure}[!t]
	\centering
	\vspace{0em}
	\includegraphics[width=1\linewidth]{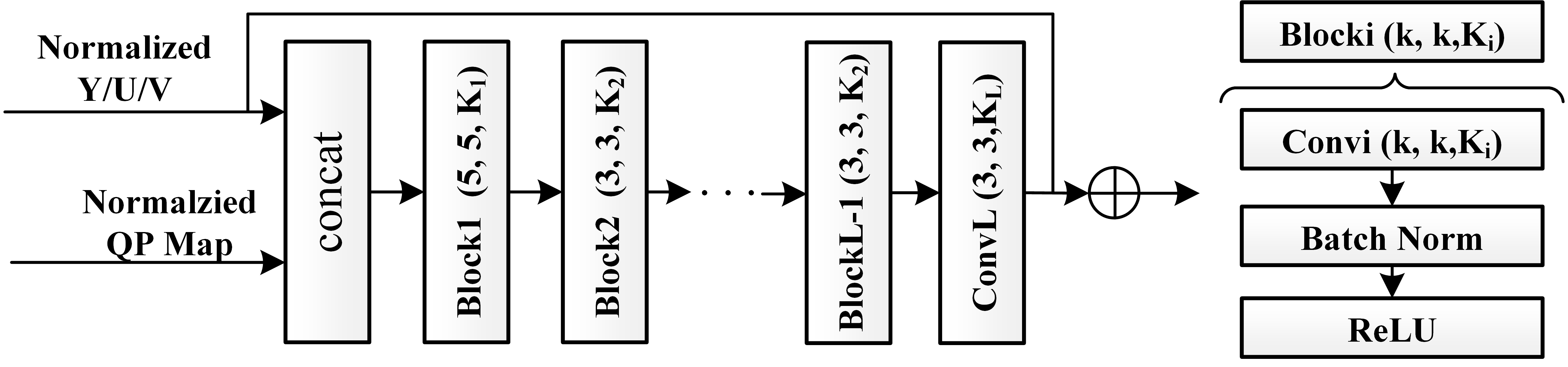}
	\caption{Network structure of CNNF, in which``Convi'' represents a convlution layer, $k$ is the kernel size and $K_L$ is the kernel number.}
	\setlength\textfloatsep{0mm}

	\label{network}
	\vspace{-1.5em}
\end{figure}

\section{The Proposed CNNF}
\label{sec:pagestyle}
\subsection{Network structure}
\label{net}
CNNF includes two inputs:  the reconstruction and QP map, which makes it possible to use a single set of parameters to adapt to reconstructions with different qualities. QP map is generated by $QPMap(x, y)=QP$, where $QP$ is the QP used for encoding, $x=0,1,...,W-1$ and $y=0,1,...,H-1$. $W$ and $H$ are the width and height of the reconstruction, respectively. Note that the reconstruction can be a decoded frame or a block.

Before fed to CNNF, both the two inputs are normalized to [ 0,1] for better convergence in training process. Each pixel in the decoded frame is divided by $1<<B-1$, in which $B$ is the bit depth and $<<$ denotes bit shift. QP map is divided by the maximum value of QP. After filtering, a corresponding de-normalization is established to obtain the final reconstruction.

In the following, a simple CNN with 8 convolution layers as shown in Fig. \ref{network} is taken as an example to make a trade-off between performance and complexity. We claim that taking QP map as a side information can also be applied to other network, e.g. \cite{VRCNN,wang2017novel, Kang2017MultimodalmultiscaleCN,yang2017decoder, jia2017spatial}. $K_L$ is set to 64. By connecting the normalized Y, U or V to summation layer, the network is regularized to learn characteristics of the residual between the decoded frame and its original one. 
\label{cnnf:network}

\begin{table}[!t]
	\vspace{0em}
	\caption{Compressed filter number for each convolution layer}
	\label{tableFilter}
	\centering
	\setlength{\tabcolsep}{2.5mm}{
		\begin{tabular}{ c | c | c | c | c | c | c | c }
			\hline
			convL & $K_1$ & $K_2$ & $K_3$ & $K_4$ & $K_5$ & $K_6$ & $K_7$ \\ \hline
			\# Filter& 45 & 54 & 58 & 48 & 51 & 40 & 31\\ \hline
	\end{tabular}}
\vspace{-1.5em}
\end{table}

\subsection{Model Compression}
To speed up, the learned model is compressed before testing. For efficient compression, loss function $Loss$ with two additional regularizers included is designed for the training process as the following
\begin{multline}
\label{regulation}
Loss=\underbrace{\frac{1}{2M}\sum_{i=1}^{M}||y^i-f_w(x^i)||^2}_{\textup{mean\space square\space error}}+\lambda_w\underbrace{\sum_{j=1}^{L}||W_j||_{g_1}}_{\textup{normal regularizer}}+\\
\underbrace{\lambda_s||S||_{g_2}+\lambda_{lda}\sum_{j=1}^{L-1}\sum_{i=1}^{L-1}||\frac{W_j}{||W_j||}-\frac{W_i}{||W_i||}||_1}_{\textup{additional\space regularizers}},
\end{multline}
in which $y^i$ and $f_w(x^i)$ are the ground truth and the filtered results of $x^i$, respectively. $W_{(\cdot)}$ is the parameter to be learned and $S$ is the scale parameter in BN layer \cite{BN}. $\lambda_w$, $\lambda_s$ and $\lambda_{lda}$ are hyper parameters. $M$ is the batch size. $L$ is the number of convolution layers. Mean square error is used as the main measurement of loss. The normal regularizer constraints the model complexity. $g_1$ and $g_2$ denotes $L_1$ or $L_2$ norm. To reduce overall combinational cases, we set $g_1=g_2$ and experimental results show that $L_2$ norm is better than $L_1$.

With the first additional regularizer, the learned scale parameters in BN layer tends to be zero. In the training process, a filter will be pruned once the absolute value of its corresponding scale parameter is small enough. The second additional regularizer, i.e. the linear discriminant analysis (LDA) item, makes the learned parameters friendly to the following low rank approximation. Then singular value decomposition (SVD) is established for low rank approximation\cite{denton2014exploiting}. After that, filters are reconstructed using a much lower basis \cite{wen2017coordinating}. 

Table \ref{tableFilter} gives the compressed filter number. It can be observed that it efficiently reduces the kernel number and the amount of parameters is reduced to 51\% of the original model. Experimental results report performance only changes about -0.08\%, -0.19\%, 0.25\% in average for Y, U and V components of class B, C, D and E on JEM 7.0 \cite{JEM7.0}.

\subsection{Dynamic Fixed Point Inference}
To ensure consistency between encoding and decoding across different platforms, DFP \cite{gysel2016hardware} operations are proposed to be used in testing. A value $V$ in dynamic fixed point is described by
\begin{equation}
V = (-1)^s\cdot{2^{-FL}}\sum_{i=0}^{B_{f}-1} 2^i {\cdot} x_i,
\end{equation}
where $B_{f}$ denotes bit width to represent the DFP value, $s$ the sign bit, $FL$ the fractional length and $x_i$ the mantissa binary bits. Each float point within model parameters and outputs is quantized and clipped to be converted to DFP.

Values in CNNF are divided into three groups: layer weights, biases and outputs. Bit width for weights $B_w$ and biases $B_b$ is set to 8 and 32, respectively. Since weights and biases quantization leads to performance loss, fine-tuning taking quantization into account is established similar to \cite{gysel2016hardware,LinTA15}. After that, the parameters are quantized to DFP. For layer outputs, the bit width is set to 16. Experimental results show that quantization of outputs leads to negligible loss.

\begin{table}[!t]
	\vspace{-0.5em}
	\caption{Estimated $FL$ for each convolution layer}
	\label{tableFL}
	\centering
	\setlength{\tabcolsep}{2.7mm}{
		\begin{tabular}{ c | c | c | c | c | c | c | c | c }
			\hline
			convL & 1 & 2 & 3 & 4 & 5 & 6 & 7 & 8\\ \hline
			$FL_{w}$ & 9 & 8 & 8 & 8 & 8 & 8 & 8 & 10\\ \hline
			$FL_{b}$ & 17 & 15 & 14 & 16 & 15 & 13 & 13 & 16\\ \hline
			$FL_{o}$ & 15 & 14 & 14 & 15 & 15 & 15 & 16 & 18\\ \hline
	\end{tabular}}
	\vspace{-1.5em}
\end{table}
Each group in the same layer shares one common $FL$, which is estimated from available training data and layer parameters. Table \ref{tableFL} gives the estimated $FL$ for each convolution layer.  $FL_{w}$, $FL_{b}$ and $FL_{o}$ denotes $FL$ of weights, biases and outputs, respectively. $FL$ for concat and summation layer are both set to 15. Since CPU and GPU do not support DFPs, they are simulated by float points similar to \cite{gysel2016hardware}.

\begin{figure*}[!t]
	
	\begin{minipage}[b]{0.25\linewidth}
		\centering
		\centerline{\includegraphics[width=5.5cm]{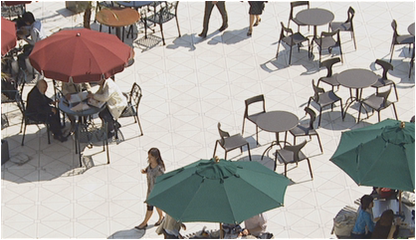}}
		\centerline{(a) The original frame}
	\end{minipage}
	\hfill
	\begin{minipage}[b]{0.25\linewidth}
		\centering
		\centerline{\includegraphics[width=5.5cm]{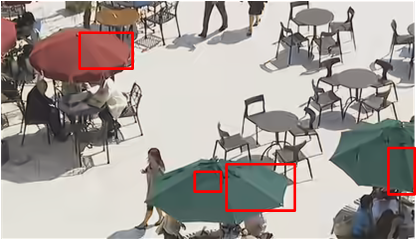}}
		\centerline{(b) The decoded frame of JEM 7.0}
	\end{minipage}
	\hfill
	\begin{minipage}[b]{0.25\linewidth}
		\centering
		\centerline{\includegraphics[width=5.5cm]{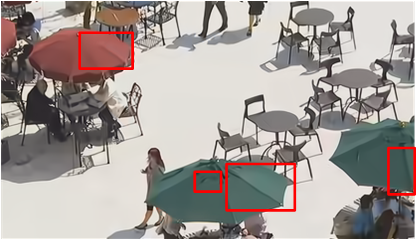}}
		\centerline{(c) The decoded frame of CNNF}
	\end{minipage}
    \vspace{-0.5em}
	\caption{Comparison of subjective quality of BQSquare encoded with QP 37. The red bounding boxes demonstrate the highlighted areas in CNNF are more enhanced than JEM 7.0.}
	\label{subQ}
	\vspace{-2em}
\end{figure*}
\section{Training Process}
\label{training}
\begin{table}[!t]
	\vspace{-0.5em}
	\caption{Evaluation environment}
	\label{environment}
	\centering
	\setlength{\tabcolsep}{2.5mm}{
		\begin{tabular}{ c | p{3cm} | p{3cm} }
			\hline
			convL & CPU+GPU & CPU \\ \hline
			CPU & Intel(R) Xeon(R)  CPU E5-2650 v4 @ 2.20GHz &Intel(R) Xeon(R)  CPU E5-2680 v4 @ 2.40GHz \\ \hline
			GPU & NVIDIA Titan Xp  with 12GB Memory & ---------------\\ \hline
			Library & cuDNN 5.1.10 & OpenBLAS 0.2.18\\ \hline
	\end{tabular}}
\vspace{-1.3em}
\end{table}

Training data used in obtaining the layer parameters of CNNF are generated from Visual genome(VG) \cite{VG}, DIV2K \cite{DIV2K} and ILSVRC2012 \cite{ILSVRC2012}. Each image is intra encoded by the QP 22, 27, 32, 37 on JEM 7.0 with BF, DF, SAO and ALF off. Then the decoded frames, including luma and chroma components, are divided into patches with 35$\times$35 size. After that 3.6 million training data are generated which includes 600 thousands luma data and 300 thousands chroma data for each QP. Finally, all the data are mixed in a random order. 

We use Caffe \cite{jia2014caffe} as the training software on a NVIDIA Tian Xp with 12GB memory GPU.  Eq. (\ref{regulation}) is used as the loss function during training. Parameters of the proposed model are initialized randomly. Batch size $M$ is set to 64. The base learning rate is set to 0.1. And $\lambda_w$, $\lambda_s$ and $\lambda_{lad}$ are set to 1e-5, 5e-8 and 3e-6, respectively. Stochastic gradient decent is used to solve the optimization with gradients clipped. The training is stopped after 32 epochs.

\section{Experimental Results}
\label{sec:experiments}
In evaluation, JEM 7.0 \cite{JEM7.0} serves as the reference software and common test condition in \cite{CTC} is used. BD-rate \cite{bjontegarrd2001calculation} is used as the measurement for comparison. Without specification, BF, SAO, DF and ALF are all turned off. Results with CPU and GPU are both tested and denoted as ``CPU+GPU'' and ``CPU'', respectively. Anchor is evaluated in the both settings, respectively. The test environment is listed in Table \ref{environment}. Since QP as a side information can be introduced to any network for loop filtering, we do not compare with other works.

\begin{table}[!t]
	\centering
	\vspace{0em}
	\caption{Performance improvement of ``QP Dependent Model'' and CNNF on luma}
	\label{QP}
	\setlength{\tabcolsep}{0.8mm}{
		\begin{tabular}{ l | c | c  c  c  c c }
			\hline
			\space & \space &\multicolumn{5}{|c}{QP Dependent Model} \\ 
			Model  & CNNF   & Best    & QP22    & QP27    & QP32    & QP37  \\ \hline
			ClassB &-2.95\% &-3.09\%  & -0.49\% & -1.67\% & 0.22\%  & 8.09\% \\ \hline
			ClassC &-4.09\% &-4.24\%  & -0.59\% & -2.44\% & -1.10\% & 6.12\% \\ \hline
			ClassD &-4.72\% &-4.90\%  & -1.02\% & -3.02\% & -1.91\% & 5.20\% \\ \hline
			ClassE &-4.65\% &-4.75\%  & -0.05\% & -2.54\% & -1.69\% & 6.03\% \\ \hline
			\bfseries{Overall} & \bfseries{-3.99\%} & \bfseries{-4.14\%} & \bfseries{-0.57\%} & \bfseries{-2.36\%} & \bfseries{-1.00\%} & \bfseries{6.49\%}\\ \hline
	\end{tabular}}
	\vspace{-1.3em}
\end{table}

\begin{table}[!t]
	\caption{Test results of AI configuration with ALF off}
	\label{AI_alf_off}
	\centering
	\setlength{\tabcolsep}{5mm}{
		\begin{tabular}{ l | c  c  c }
			\hline
			\space & Y & U & V \\ \hline
			ClassA1 & -1.57\% & -4.74\% & -4.03\% \\ \hline
			ClassA2 & -2.36\% & -5.72\% & -6.07\% \\ \hline
			ClassB & -2.71\% & -4.58\% & -5.99\% \\ \hline
			ClassC & -3.70\% & -6.21\% & -8.21\% \\ \hline
			ClassD & -4.07\% & -5.29\% & -7.98\% \\ \hline
			ClassE & -3.97\% & -5.64\% & -4.81\% \\ \hline
			\bfseries{Overall} & \bfseries{-3.14\%} & \bfseries{-5.21\%} & \bfseries{-6.28\%} \\ \hline
	\end{tabular}}
	\vspace{-1.5em}
\end{table}
First, effectiveness of the QP independent model is evaluated, in which model compression and DFP inference are not used. The above network without QP map is used for comparison and denoted as ``QP dependent Model''. A model is trained for QP 22, 27, 32 and 37 and denoted as ``QP22'', ``QP27'', ``QP32'' and ``QP37'', respectively. The encoded data generated by its corresponding QP in Section \ref{training} is used as training data. In the test, decoded frames with all QPs are filtered by each model, respectively. The results that decoded frames with different QP using its corresponding model are also tested and denoted as ``Best''.

Table \ref{QP} gives the test results above. Compared with ``QP dependent Model'', CNNF shows large gains when using a single model for all QPs. And CNNF even shows comparative gains over ``Best''. 

Table \ref{AI_alf_off} shows the results of CNNF with all intra (AI) configuration. It can achieve 3.14\%, 5.21\% and 6.28\% BD-rate savings for luma and both chroma components. The subjective quality is given in Fig. \ref{subQ}. It can be observed that subjective quality is enhanced, especially in the edge area.

Table \ref{AI} and Table \ref{RA} show results on video coding. CNNF is only applied to intra frames. Due to inter dependency within ALF, it is not replaced. For B and P frames, filters are configured the same as JEM 7.0. 3.57\%, 6.17\% and 7.06\% average gains are observed with AI configuration. Though only applied to intra frames, CNNF achieves 1.23\%, 3.65\% and 3.88\% gains with RA configuration. 

Table \ref{AI} and Table \ref{RA} also compare the encoding time (EncT) and decoding time (DecT), which are measured by the ratio of time consuming of the proposed scheme to that of JEM 7.0 during encoding and decoding, respectively. With GPU, the EncT decreases and DecT increases a little. Even when testing with CPU, the EncT only increases a little. Though DecT is extremely high on CPU, we do believe that with the development of deep learning specific hardware it will not be a problem.

\begin{table}[!t]
	\caption{Test results of AI configuration with ALF on}
	\label{AI}
	\centering
	\setlength{\tabcolsep}{0.2mm}{
	\begin{tabular}{ l | c  c  c | c  c | c  c }
		\hline
		 \space &\space  &\space  &\space  & \multicolumn{2}{|c|}{CPU+GPU} & \multicolumn{2}{|c}{CPU}\\ 
		  \space & Y & U & V & EncT & DecT & EncT & DecT\\ \hline
		ClassA1 & -2.26\% & -6.21\% & -5.05\% & 93\% & 157\% & 109\% & 15360\%\\ \hline
		ClassA2 & -3.58\% & -6.33\% & -7.02\% & 92\% & 158\% & 112\% & 16312\%\\ \hline
		ClassB & -3.08\% & -5.06\% & -6.27\% & 94\% & 148\% & 108\% & 15360\%\\ \hline
		ClassC & -3.88\% & -6.98\% & -9.11\% & 94\% & 158\% & 103\% & 11139\%\\ \hline
		ClassD & -4.13\% & -5.63\% & -8.20\% & 94\% & 214\% & 102\% & 7256\%\\ \hline
		ClassE & -4.93\% & -7.41\% & -6.88\% & 94\% & 169\% & 111\% & 15441\%\\ \hline
	    \bfseries{Overall} & \bfseries{-3.57\%} & \bfseries{-6.17\%} & \bfseries{-7.06\%} & \bfseries{93\%} & \bfseries{157\%} & \bfseries{109\%} & \bfseries{12887\%}\\ \hline
	\end{tabular}}
\vspace{-1.2em}
\end{table}
\begin{table}[!t]
	\caption{Test results of RA configuration with ALF on}
	\label{RA}
	\centering
	\setlength{\tabcolsep}{2mm}{
		\begin{tabular}{ l | c  c  c | c  c }
			\hline
			\space &\space  &\space  &\space  & \multicolumn{2}{|c}{CPU} \\ 
			\space & Y & U & V & EncT & DecT \\ \hline
			ClassA1 & -0.39\% & -1.96\% & -1.93\% & 99\% & 275\%  \\ \hline
			ClassA2 & -1.76\% & -3.70\% & -4.29\% & 99\% & 303\%  \\ \hline
			ClassB & -1.46\% & -4.65\% & -4.14\% & 99\% & 339\%  \\ \hline
			ClassC & -1.28\% & -4.40\% & -4.75\% & 99\% & 289\% \\ \hline
			ClassD & -1.22\% & -3.28\% & -4.20\% & 99\% & 219\% \\ \hline
			\bfseries{Overall} & \bfseries{-1.23\%} & \bfseries{-3.65\%} & \bfseries{-3.88\%} & \bfseries{99\%} & \bfseries{284\%} \\ \hline
	\end{tabular}}
    \vspace{-1.7em}
\end{table}



\section{Conclusion}
\label{sec:conclusion}
This paper proposes a practical CNN as the loop filter for intra frames. It uses a single model with low redundancy for loop filtering, which can adapt to reconstructions with different qualities. Besides, it solves the problem of mismatched encoding and decoding results across various platforms. Comparing with JEM 7.0, the proposed CNNF achieves large gains though only applied for intra frames. More gains will be expected to extend it to B and P frames in future work.

\bibliographystyle{IEEEbib}

\end{document}